\documentclass{ws-ijmpa}

\newcommand{\by}{{\bf y}}
\newcommand{\br}{{\bf r}}
\newcommand{\bl} {\mbox{\boldmath $l$}}
\newcommand{\eqntimes}{\mbox{} \times}

\begin{document}

\markboth{P.R. Page}{Hybrid and Conventional Baryons}

\catchline{}{}{}{}{}

\title{HYBRID AND CONVENTIONAL BARYONS IN THE FLUX-TUBE AND QUARK MODELS}

\author{PHILIP R. PAGE}
\address{Theoretical Division, MS B283, Los Alamos National Laboratory,
Los Alamos, NM 87545, USA. E-Mail: prp@lanl.gov}


\maketitle


\begin{abstract}

The status of conventional baryon flux-tubes and hybrid baryons
is reviewed. Recent surprises are that a model prediction indicates
that hybrid baryons are very weakly produced in
glue-rich $\Psi$ decays, and an analysis of electro-production data
concludes that the Roper resonance is not a hybrid baryon.
The baryon decay flux-tube overlap has been calculated in the
flux-tube model, and is discussed here. The behavior of the overlap 
follows na\"{\i}ve expectations.
 
\end{abstract}



\section{Update on previous reviews}

Comprehensive reviews on baryon flux-tubes and hybrid baryons 
are available.~\cite{bar00,page02} Here developments since
the last review in 2002 are summarized.

\begin{figure}
\centerline{\psfig{file=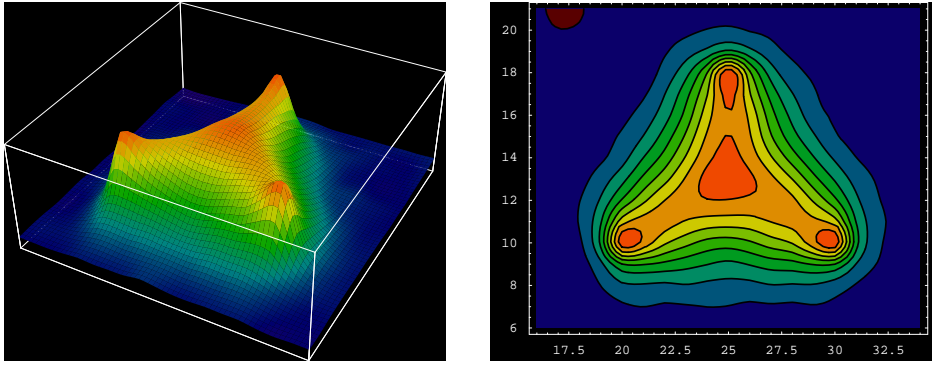,width=1.02\linewidth}}
\vspace*{-5pt}
\caption{\label{ich}Side view and top-down view of the flux action density.}
\end{figure}

{\it Baryon flux-tube:} The consensus in quenched lattice QCD is that the
baryon potential at
sufficient distances is best described by the sum of a Coulomb, Y-shaped
confinement, and constant term.~\cite{takahashi} This statement
is independent of the lattice operator, and hence
physical.~\cite{woloshyn} The difference with Abelian projected full 
lattice QCD is small.~\cite{ichie} The baryon flux action 
density~\cite{ichie}
is displayed in Fig.~\ref{ich}.  The density clearly peaks at the
three quarks and the junction. However, the detailed contour lines,
as well as the Y-shape, are lattice operator
dependent.~\cite{woloshyn} Abelian projected full lattice QCD
shows~\cite{ichie} that cromo-electic flux flows in the same
directions as the flux-tubes, with solenoidal currents circulating
around them. This supports~\cite{ichie} the idea that QCD is a dual 
superconductor.

\begin{figure}
\centerline{\psfig{file=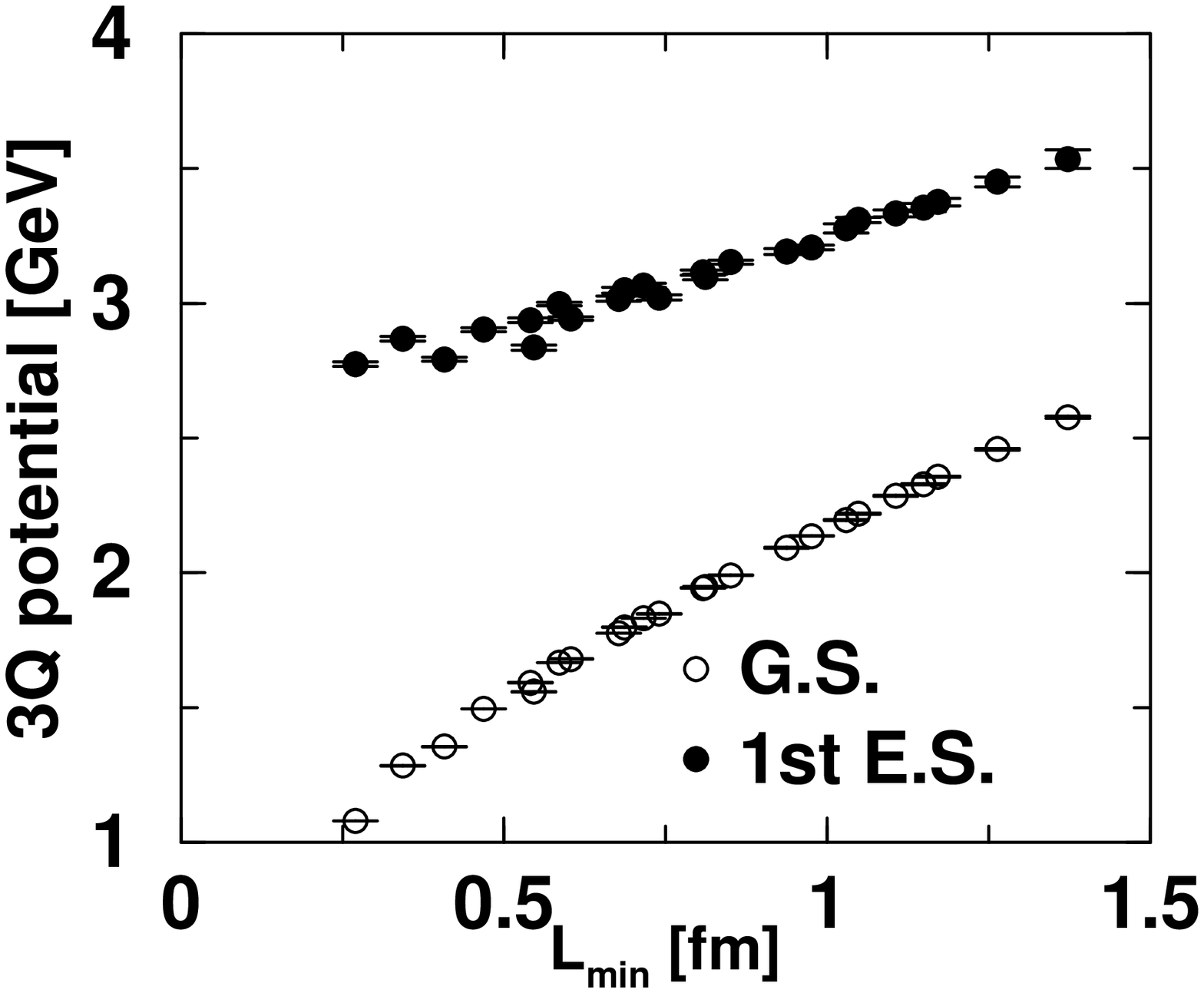,width=.55\linewidth}
\psfig{file=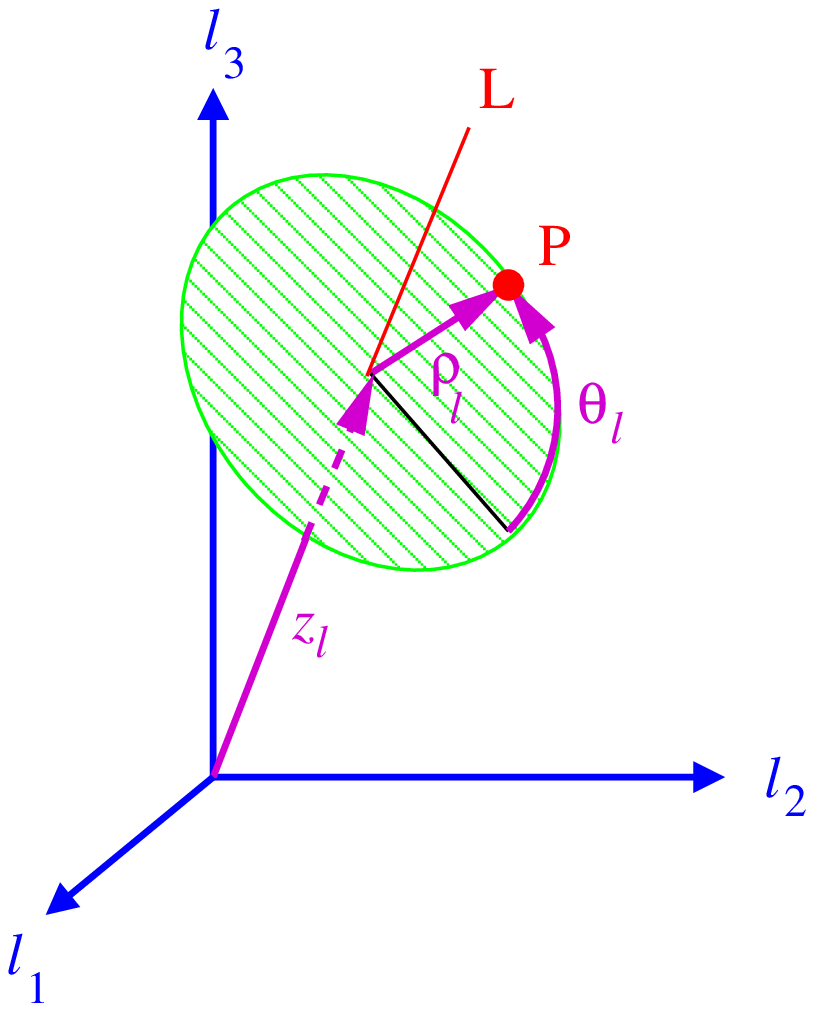,width=0.40\linewidth}}
\vspace*{-4pt}
\caption{\label{tak}Left is the baryon and lowest hybrid baryon potential 
as a function of ${\mbox{L}}_{\mbox{min}}\equiv l_1+l_2+l_3$. Here 
$l_i$ is the distance between quark $i$ and the junction equilibrium 
position, shown in Fig.~\protect\ref{def}.
Right is a cylindrical coordinate system where the point P $=(l_1,l_2,l_3)$ is
rewritten in terms
of $\rho_l,\theta_l$ and $z_l$. The line $l_1=l_2=l_3$ is denoted by L.
} 
\end{figure}

{\it Hybrid baryon:} The first quenched lattice QCD calculation of the
low-lying hybrid baryon potential has been performed.~\cite{taka} The
lowest gluonic excitation energy (Fig.~\ref{tak}) is consistently
greater than 1 GeV. This is several hundred MeV larger than the
flux-tube model prediction~\cite{capstick}, which already gives the
highest mass prediction for the lightest $N\frac{1}{2}^+$ hybrid
baryon of all the models~\cite{bar00,page02}. If the lattice calculation
is correct, this hybrid baryon will likely have a mass above 2 GeV, which
would mean that the hybrid baryon spectrum starts at a very high mass.
In contrast to this, QCD sum rule predictions for hybrid baryon masses
remain low: Most recently, the $\Lambda\frac{1}{2}^+$ is predicted~\cite{kiss}
at $1.6(2)$ GeV. A possible reason why the lattice calculation is high is that
the effect of higher excited states has not been eliminated, although
it is stated~\cite{taka} that this is not the case.

{\it Hybrid baryon production in $\Psi$ decay}, a glue-rich process, 
has long been thought to be
promising.~\cite{page02} Particularly, BES at BEPC can look for
$\Psi\rightarrow \mbox {hybrid} \;\bar{N}$. Recently, several high
mass peaks have been reported~\cite{bes}: a $J^P=\frac{1}{2}^+\;
2\sigma$ peak at $1834^{+46}_{-55}$ MeV seen in $\Psi\rightarrow
p\bar{p}\eta$, and a $2068\pm 3^{+15}_{-40}$ MeV peak seen in
$\Psi\rightarrow N\bar{N}\pi$.  A recent model calculation~\cite{ping}
finds, in contrast to na\"{\i}ve expectations, that hybrids are very
weakly produced relative to radially excited baryons in $\Psi$ decay. 
This is due to a low probability of the gluon coming from the $\Psi$
to fold in to the $qqqg$ hybrid wavefunction, and color factors in the
hybrid baryon wave function. In fact, it is suggested that
$\Psi$ decay is a filter in favor of conventional baryons.~\cite{ping}

{\it Hybrid baryon electro-production:} A recent analysis of new
Jefferson Lab data finds~\cite{volker} that the amplitude for Roper
resonance ($N \frac{1}{2}^+(1440)$) production via longitudinal
photons is substantial, inconsistent with a model prediction that it
should be zero. The Roper hence does not fit well as a hybrid baryon.

\begin{figure}
\centerline{\psfig{file=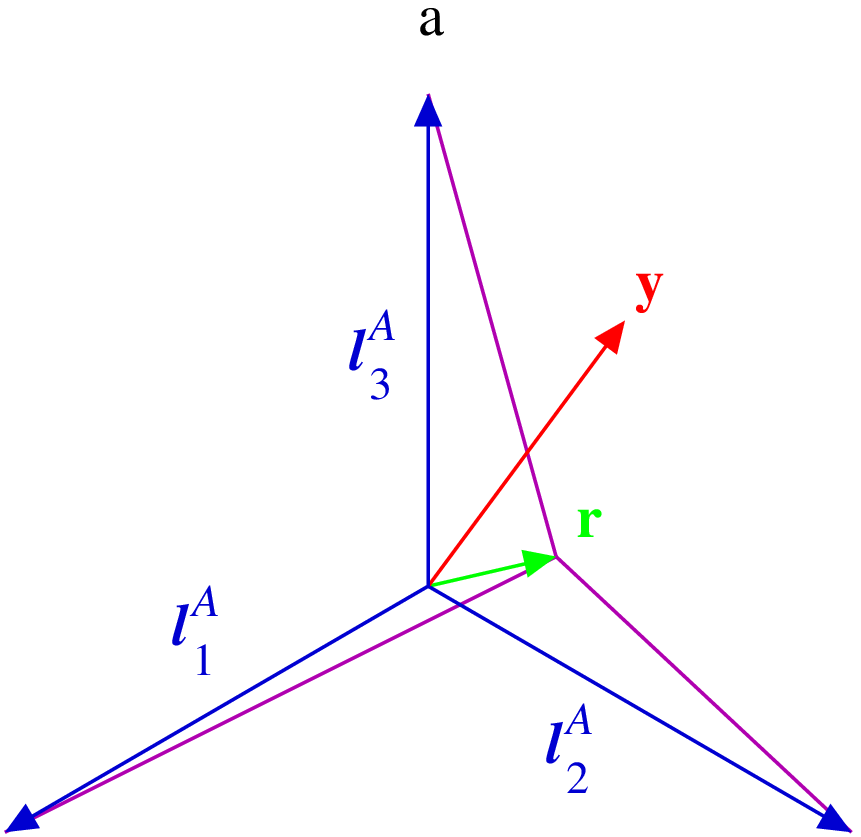,width=0.44\linewidth}
\psfig{file=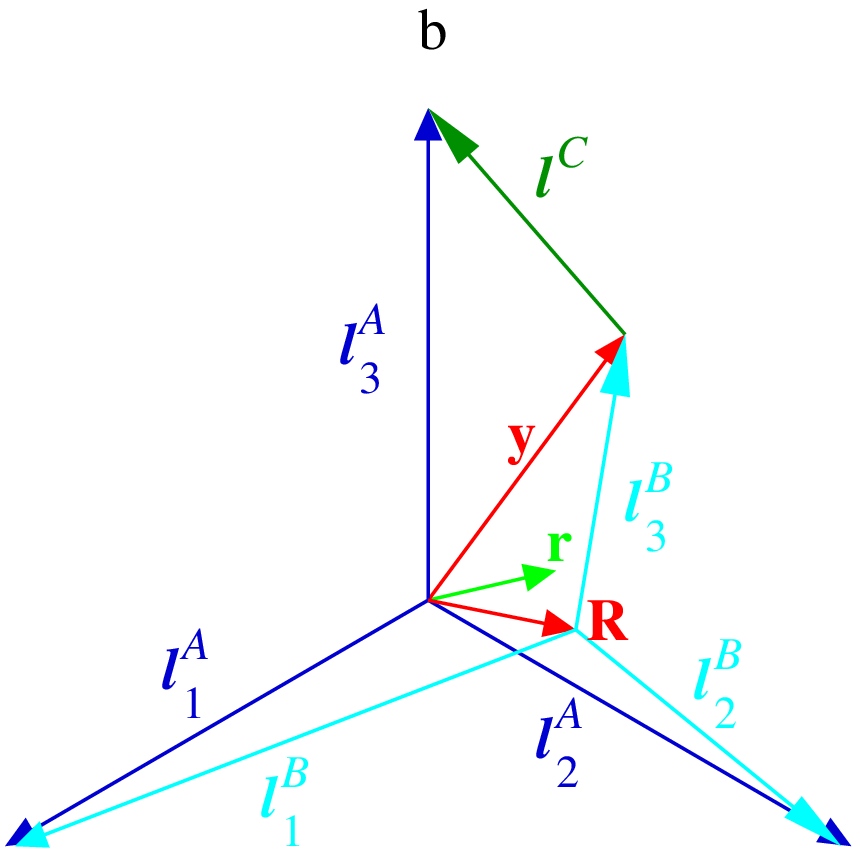,width=0.44\linewidth}}
\vspace*{-4pt}
\caption{\label{def} Coordinate system described in the text.
The length $l_i$ used in the text equals $|\bl^A_i|$.
The junction equilibrium position of the
final baryon $B$ is denoted by $\bf R$.} 
\end{figure}

\section{Baryon decay flux-tube overlap}

When a baryon decays via OZI allowed quark-antiquark pair creation
to a baryon and a meson, it has not hitherto been
considered what happens to the flux-tubes. There is certainly 
an effect, as the pair is created from gluons. Here we report on the
first preliminary calculation. Such a calculation exists for meson
decay to two mesons.~\cite{perantonis,kokoski}
The structure of the decay can be investigated by fixing the initial 
quarks in the baryon, and evaluating the overlap of the initial flux
configuration with the final baryon and mesons, for a fixed quark-antiquark
pair created from the initial flux-tube: the ``flux-tube overlap''.
This quantity enters in models of baryon decay where it is traditionally
assumed to be a constant. This is equivalent to taking the
Y-shaped flux structure to have no effect on decay. 
In such a picture 
the main effect of confinement on the decay is that it 
specifies the decay operator for quark-antiquark pair creation, 
e.g. the $^3P_0$ decay operator~\cite{kokoski}. 
This operator is typically taken to act locally with no knowledge 
of the global Y-shaped structure. Here the concern is not with the
local character of decay (the decay operator), but with imposing
constraints from Y-shaped confinement (the flux-tube overlap).
The baryon decay flux overlap

\begin{eqnarray}\lefteqn{\hspace{-3.5cm} \gamma(l_i,\by)\equiv 
\sum_{N^C=1}^{N^A_3-2}\int d^3r\; (\prod_{i=1}^3 
\prod_{j=1}^{N^A_i} d^2 y_j^{i\; A})\; 
\delta^2 (\by_{N^C+1}^{3\; A}
-(\by-\bl_3^A-\frac{N^C+1}{N^A_3+1}(\br-\bl_3^A)))\; 
\nonumber} \\ & & 
\hspace{-0.9cm} \eqntimes\psi^A(\br,\by_{j}^{i\; A})\;
\;\psi^{B\: \ast}(\br^B,\by_{j}^{i\; B})\;
\;\psi^{C\: \ast}(\by_{j}^{C}) \label{defeq}
\end{eqnarray}
measures the overlap of the flux-tube wavefunction of initial baryon
A, with those of final baryon B and meson C, assuming that pair
creation takes place at position $\by$. The overlap is evaluated
analytically in the flux-tube model of Isgur and
Paton~\cite{capstick,paton85} as was done for the meson
case~\cite{perantonis,kokoski}. In this model the flux is represented by
oscillating beads, and for the baryon the flux-tubes meet at a
junction. The coordinates of the quarks and junction
are displayed in Fig.~\ref{def}(b). The
position of the junction of the initial baryon, 
measured from its equilibrium position,
is denoted by $\br$. In Eq.~\ref{defeq} the position of 
bead $j$ associated with quark $i$ is denoted 
by $\by^i_j$. For the calculation shown, the constituent quark
masses we taken to be 330 MeV, the string tension that of mesons
($0.18$ GeV$^2\approx 1$ GeV/fm),~\cite{capstick,paton85} 
and all the bead masses equal. Bead masses are fixed
by the condition that their sum equals the potential energy in
the flux-tubes when the system is in 
equilibrium. The flux-tube overlap $\gamma$ depends on
the six variables $\by$ and $l_i$, shown in Fig.~\ref{def}(a). There
is also dependence on the number of beads associated with each
flux-tube in the initial baryon ($N^A_i$). If pair creation can take
place on more than one bead, $\gamma$ is the sum of the possible
overlaps. (Hence the sum over the number of beads on the final meson
$N^C$ in Eq.~\ref{defeq}.)
Without loss of generality pair creation associated with the
flux-tube of the third quark is studied. The components of $\by$ perpendicular
to the vector $\bl^A_3$ are $y_1$ (in the three-quark plane) and $y_3$
(perpendicular to the three-quark plane). The component of $\by$ along
$\bl^A_3$ is $y_2$. When any variables are fixed in the Figures, the
following choices are made, unless indicated otherwise: 
$y_1=y_3=0$, $y_2 = 1.25$ GeV$^{-1} =
0.25$ fm, $l_i = 2.5$ GeV$^{-1} = 0.5$ fm and $N^A_i=3$.

\begin{figure}
\centerline{\psfig{file=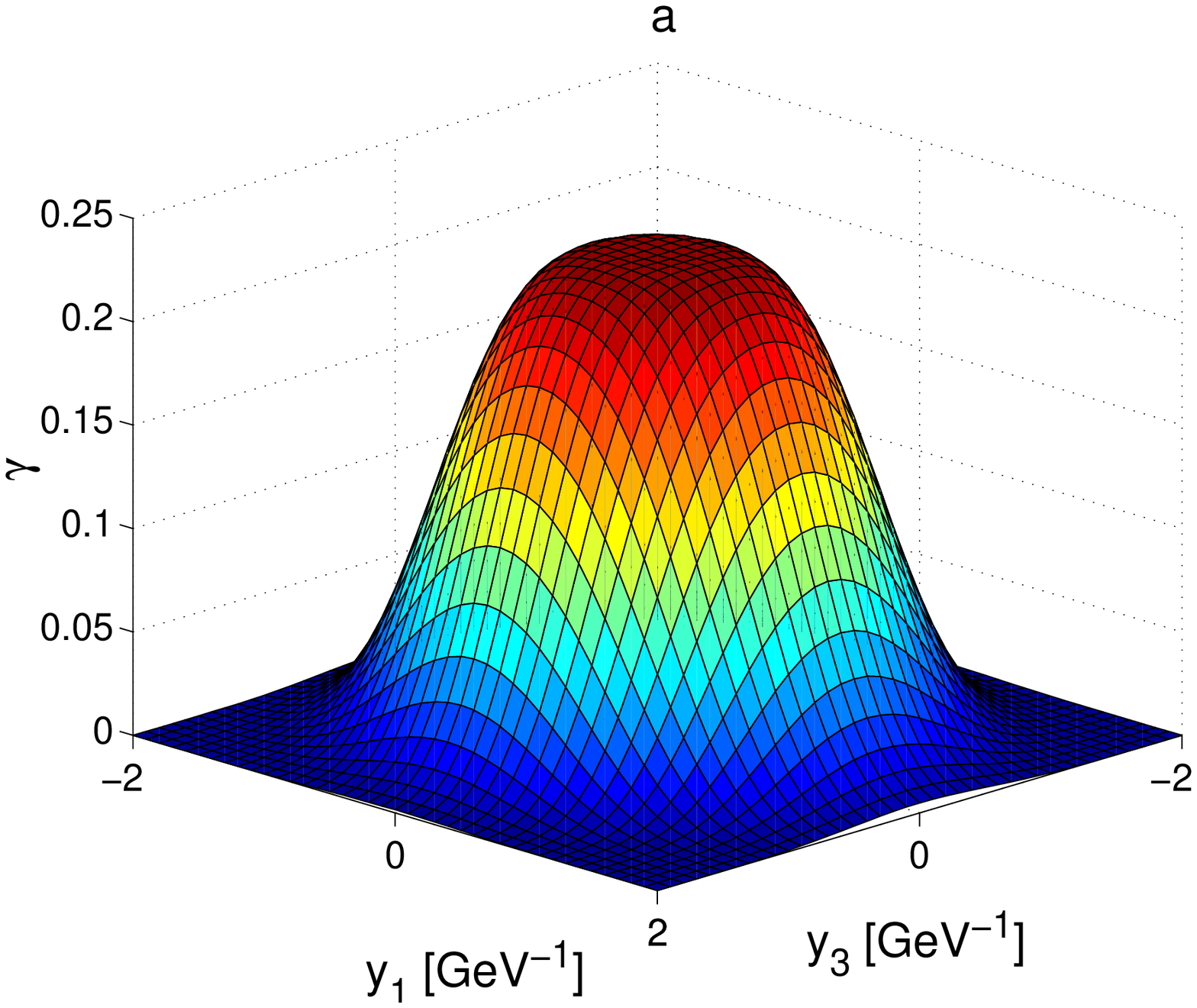,width=0.49\linewidth},
\psfig{file=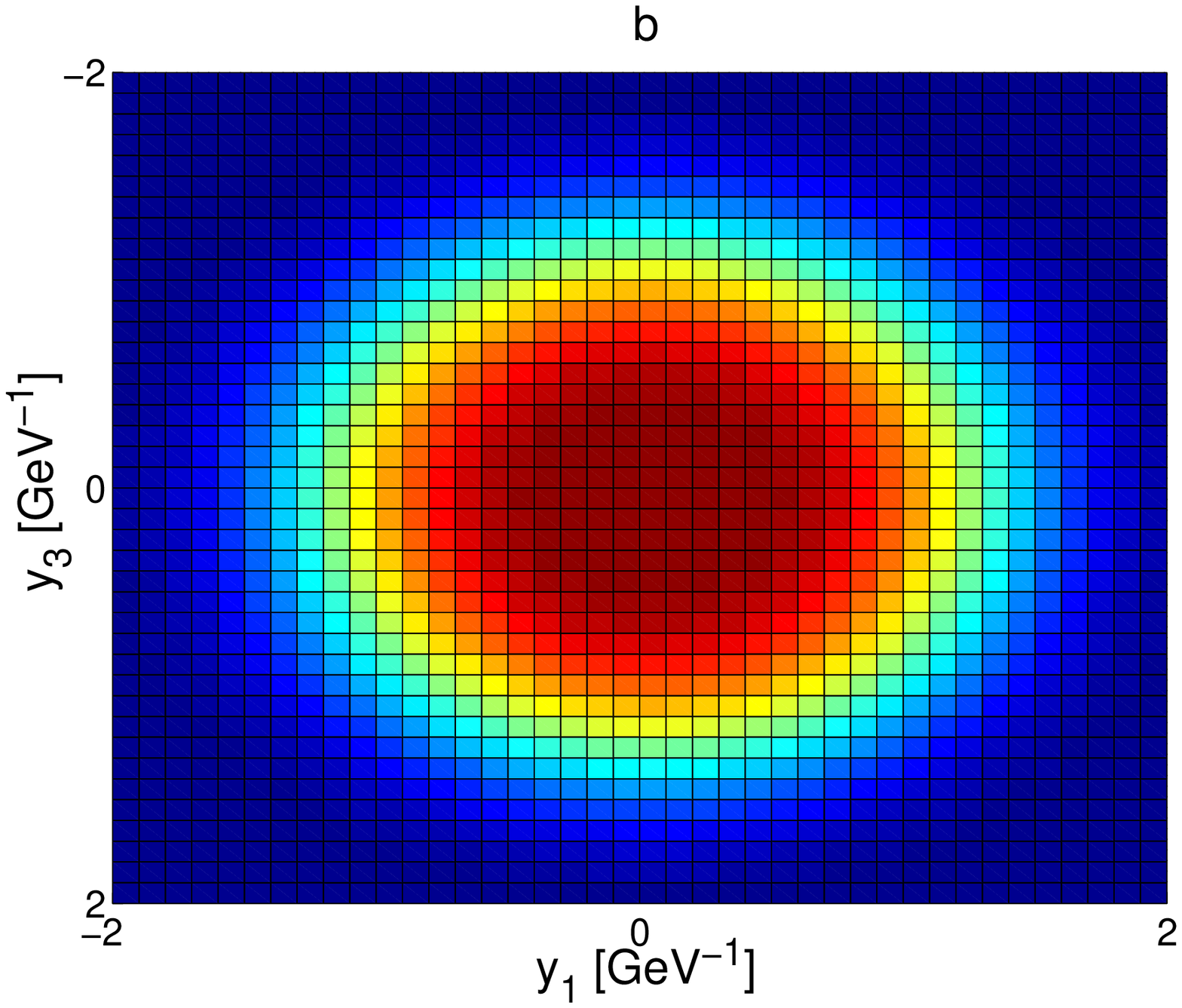,width=0.49\linewidth}}
\vspace*{-4pt}
\caption{\label{gauss} (a) Side view and (b) top-down view of the baryon 
decay flux-tube overlap $\gamma$ in GeV. It is  
invariant under $y_3\rightarrow -y_3$ (there is nothing
special about whether pair creation is above or below the 
plane of the three quarks), but the
invariance under $y_1\rightarrow -y_1$ only ensues because $l_1=l_2$.
} 
\end{figure}

\begin{figure}
\centerline{\psfig{file=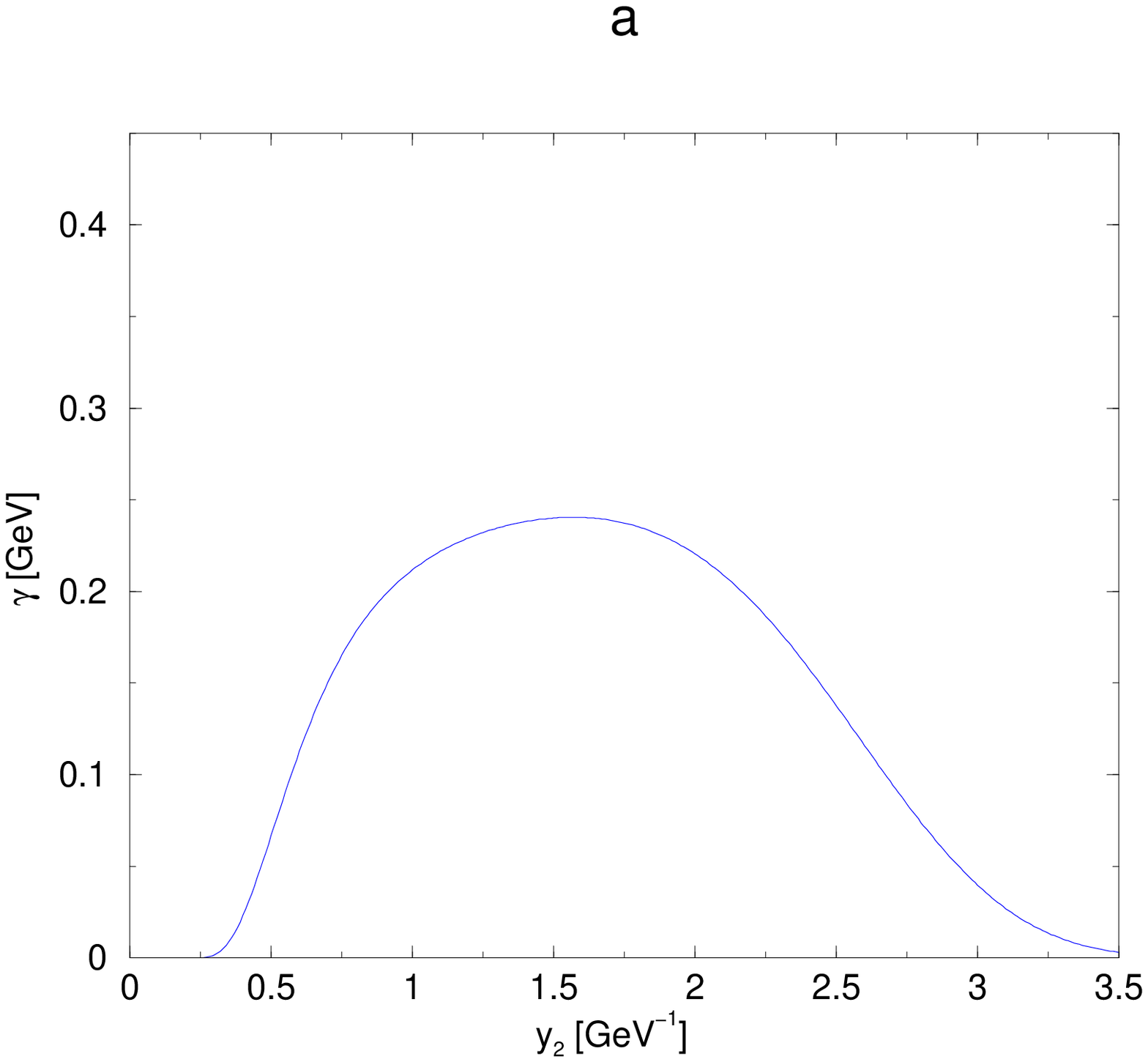,width=0.53\linewidth},
\psfig{file=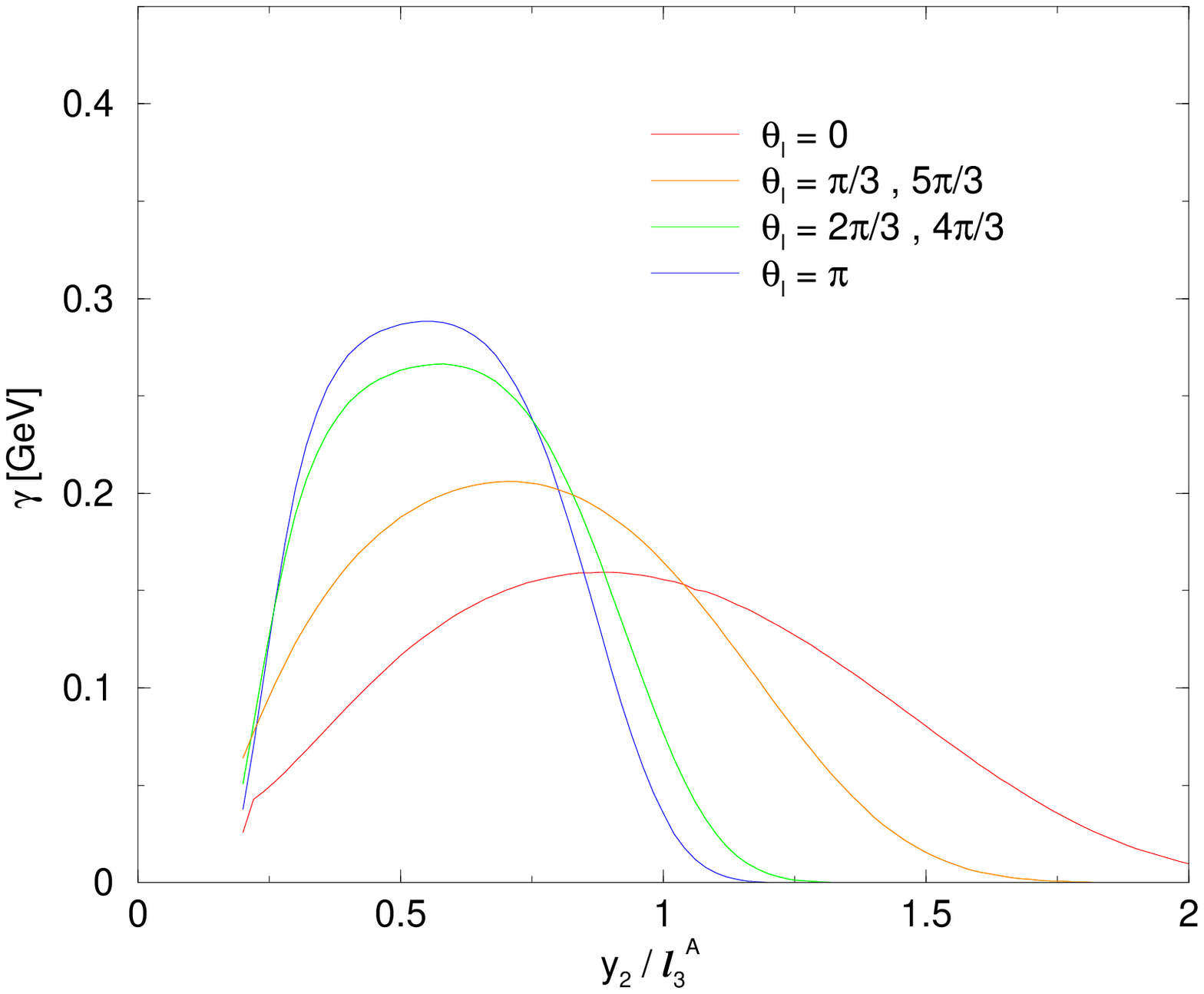,width=0.49\linewidth}}
\vspace*{-4pt}
\caption{\label{bump} The variation of $\gamma$ with $l_i$ shown in
(b) is achieved by redefining $l_i$ in terms of $\rho_l,\theta_l$ and
$z_l$ defined in Fig.~\protect\ref{tak}, and plotting the variation as
a function of $\theta_l$. The calculations are for $\rho_l =
\rho_{\mbox{\small max}}/2$ ($\rho_{\mbox{\small max}}=z_l/\sqrt{2}$)
and $z_l=\sqrt{3} \times2.5$ GeV$^{-1}$.}
\end{figure}

\begin{figure}
\vspace{-.3cm}
\centerline{\psfig{file=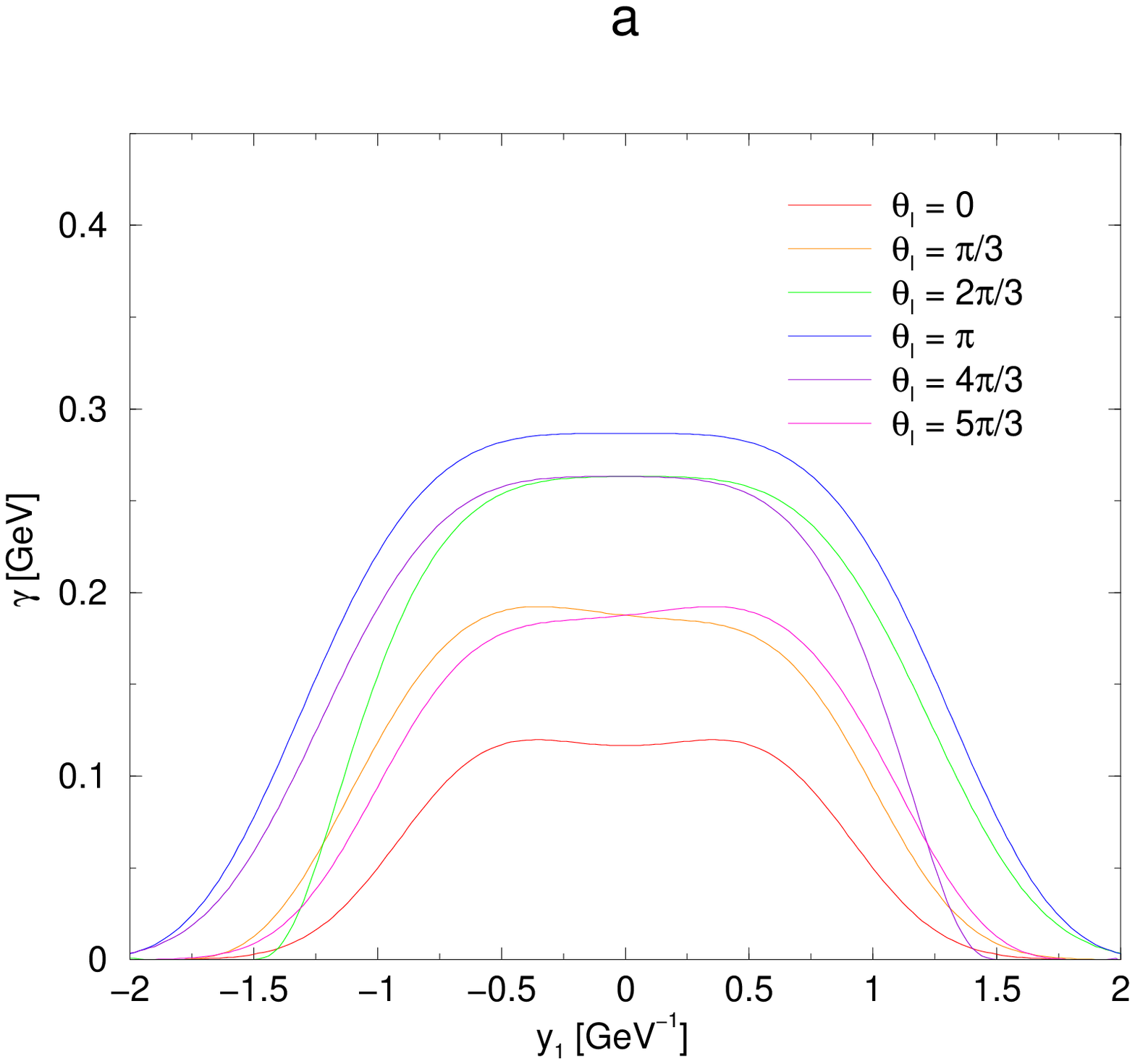,width=0.49\linewidth},\psfig{file=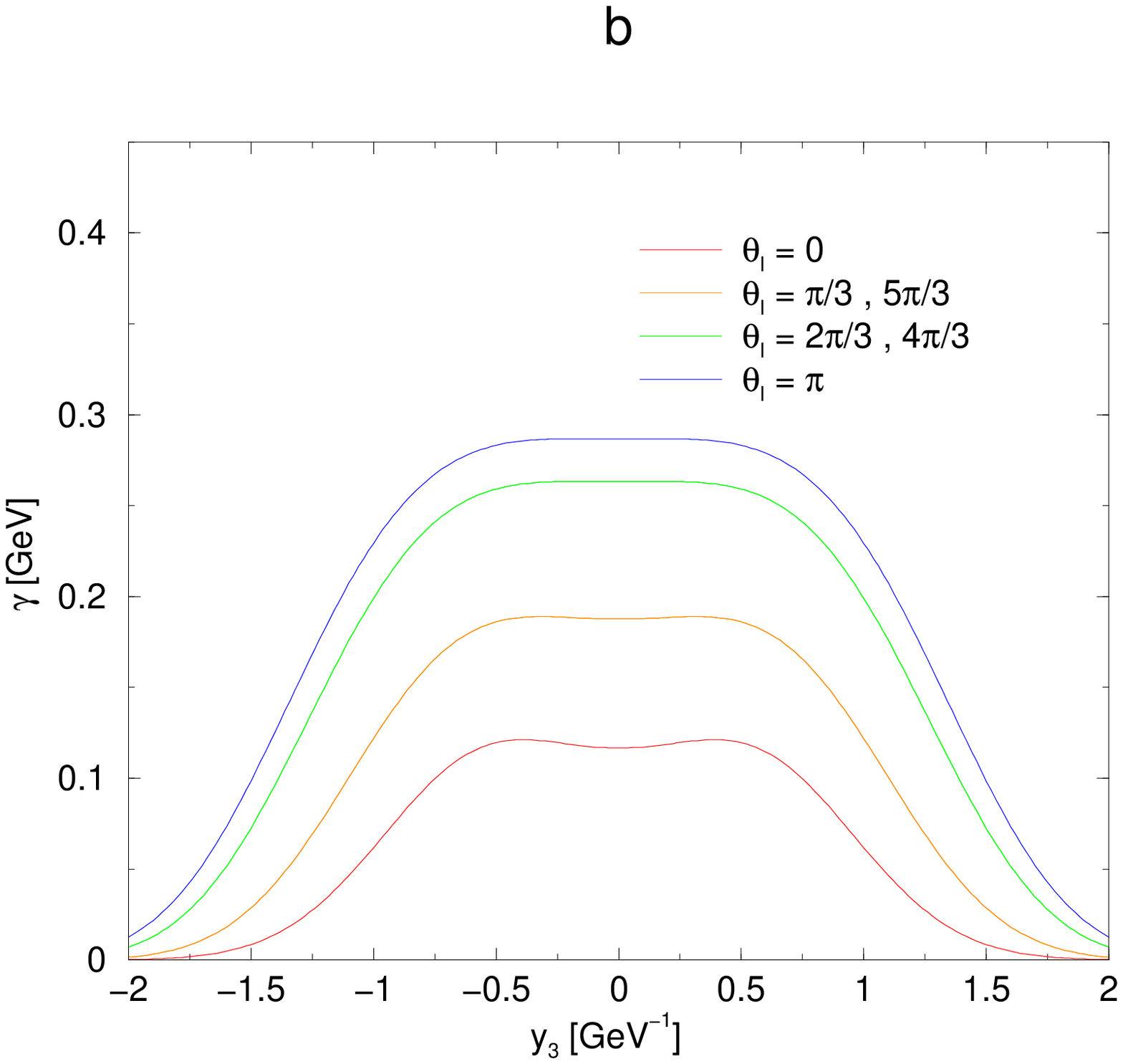,width=0.49\linewidth}}
\vspace*{-5pt}
\caption{\label{ldep} Conventions as in Fig.~\protect\ref{bump}(b),
except that $y_2=l_3/2$.}
\end{figure}

The dependence of $\gamma$ on the pair creation position $\by$ is as
follows. The dependence on $y_1$ and $y_3$ is shown in
Fig.~\ref{gauss}. A clear Gaussian fall-off with flattened top with 
a diameter $\sim 0.4$ fm think can be seen. This follows closely what would be
expected for a thick flux-tube which falls off away from $\bl^A_3$ as
predicted by the flux-tube model (see Fig. 2 of Ref.~\cite{perantonis}).
The Gaussian
fall-off with flattened top continues to be observed as $y_2$ is
varied. Fig.~\ref{bump}(a) shows the dependence on $y_2$. The
asymmetric shape is because pair creation can take place beyond the
position of the third quark ($y_2=l_3=2.5$ GeV$^{-1}$). Vanishing pair
creation at $\by=$ {\bf 0} is an artifact of the small-oscillations
approximation~\cite{capstick,perantonis,kokoski,paton85} 
made in this calculation, and is {\it
not} a prediction of the flux-tube model.
The dependence of $\gamma$ on quark position $l_i$ variations is
shown in Figs.~\ref{bump}(b) and \ref{ldep}. The extremes of the
variation are consistently 
(a) smallest $\gamma$ when $l_3 < l_1 = l_2$ (red line
$\theta_l=0$), and (b) largest $\gamma$ when $l_3 > l_1 = l_2$ (blue
line $\theta_l=\pi$). These extremes are easily understood
intuitively: When $l_3$ is small / large, there is less / more
opportunity for pair creation associated with the third quark.
Fig.~\ref{bump}(b) shows that pair creation becomes
more confined to the region inside the third quark 
($0 \leq y_2 \leq l_3$) as $l_3$ increases.
There is also a dependence on the number of beads
$N^A_i$. Calculations were also performed for $N^A_i=4$. The number of
beads determine the inter-bead spacing, a regularization parameter,
which is unphysical. The qualitative features of the dependence of
$\gamma$ on $\by$ and $l_i$ discussed above are independent of
$N^A_i$, and hence physical. They are also without exception what one
would have expected before the calculation was performed. The overall
magnitude of decay is strongly dependent on $N^A_i$, and hence
unphysical. This is as expected, since the magnitude of decay is
normalized to phenomenology in pair creation models.

\section{Conclusions}

{\it Baryon:} At sufficient distances the 
potential is Y-shaped with a clear junction. The decay flux-tube overlap
has been evaluated in the flux-tube model, and follows na\"{\i}ve
expectations.  Y-shaped baryon confinement can henceforth be
incorporated in models of decay. {\it Hybrid baryon:} The only known
way to study them rigorously is via excited adiabatic
potentials.~\cite{page02} The low--lying state in models is
$N\frac{1}{2}^+$ with mass 1.5 -- 1.8 GeV.~\cite{page02} If recent
lattice QCD calculations are correct, this mass is most likely above 2
GeV. A recent model calculation shows that glue-rich $\Psi$ decay very
weakly produces hybrid baryons.  The Roper resonance does not appear
to be a hybrid baryon from a recent analysis of electro-production via
longitudinal photons.

\vspace{-.15cm}
\section*{Acknowledgments}

Collaboration with N.F. Black (University of Tennessee, Knoxville)
on the baryon decay flux-tube overlap, and hospitality at MENU2004, 
are acknowledged.


\end{document}